# Effect of Ion Migration Induced Electrode Degradation on the Operational Stability of Perovskite Solar Cells

*Boris Rivkin,[1,2] Paul Fassl,[1,2] Qing Sun,[1,2] Alexander D. Taylor,[1,2] Zhuoying Chen[3] and Yana Vaynzof[1,2]\**

[1] Kirchhoff Institute for Physics, Im Neuenheimer Feld 227, 69120, Germany

[2] Centre for Advanced Materials, Im Neuenheimer Feld 225, 69120, Germany

[3] Laboratoire de Physique et d'Etude des Matériaux (LPEM), ESPCI Paris, PSL Research University, CNRS, Sorbonne Université, 10 Rue Vauquelin, 75005 Paris, France

*Correspondence to: vaynzof@uni-heidelberg.de

ABSTRACT

Perovskite-based solar cells are promising due to their rapidly improving efficiencies, but suffer from instability issues. Recently it has been claimed that one of the key contributors to the instability of perovskite solar cells is ion migration induced electrode degradation, which can be avoided by incorporating inorganic hole blocking layers (HBL) in the device architecture. In this work, we investigate the operational environmental stability of methylammonium lead iodide (MAPbI$_3$) perovskite solar cells that contain either an inorganic or organic HBL, with only the former effectively blocking ions from migrating to the metal electrode. This is confirmed by X-ray photoemission spectroscopy measured on electrodes of degraded devices, where only electrodes of devices with an organic HBL show a significant iodine signal. Despite this, we show that when these devices are degraded under realistic operational conditions (i.e. constant illumination in a variety of atmospheric conditions), both types of devices exhibit nearly identical degradation behavior. These results demonstrate that contrary to prior suggestions, ion-induced electrode degradation is not the dominant factor in perovskite environmental instability under operational conditions.

INTRODUCTION

The favorable electronic and optical properties of hybrid lead-halide perovskites have enabled the remarkable performance increase in solar cells based on such materials, which are currently reaching an efficiency of 22.7%.[1] Device stability, however, remains a major factor impeding their commercialization, where lifetimes of more than 25 years are required.[2] Previous studies have investigated the degradation of device performance due to environmental factors such as oxygen,[3–5] moisture,[6–9] and heat[10,11] and explanatory models have been proposed for some scenarios.[12,13] At the same time, device degradation under inert (nitrogen atmosphere) operating conditions is a widely observed but not conclusively explained phenomenon.[14,15] In one popular model, it is claimed that device degradation originates predominantly from metal electrode corrosion: mobile halide ions, such as iodide, migrate through the active material and the adjacent electron extraction layer towards the metal cathodes, such as silver.[16–20] This could then give rise to the formation of insulating silver iodine,[21,22] which would inhibit the extraction of charges, increase series resistance, and enable the formation of an undesirable dipole interface layer.[17,23] This hypothesis is often accompanied by the claim that a buffer layer of a dense electron transporting material, such as ZnO, TiO$_x$ or SnO$_2$, could serve as an "ion-blocking layer" and prohibit the migration of ions and thus improve the stability of perovskite solar cells.[24–27] While such studies have demonstrated increased device shelf life (dark storage) stability, few studies featuring full device degradation, under constant illumination and both inert and non-inert atmospheres, as well as employing rigorous compositional analysis to conclusively prove either of these two claims have been presented to date.[27]

In this work, we compare the environmental stability of methylammonium lead iodide (MAPbI$_3$) perovskite solar cells that contain inorganic ZnO nanoparticle-based HBL against reference devices that employ the commonly used bathocuproine (BCP) as the HBL. While both types of devices show a similar initial photovoltaic performance, the inorganic HBL effectively blocks mobile iodide ions from reaching and reacting with the metal electrode. This is not the case for devices with an organic HBL, in which a significant amount of iodine is detected by X-ray photoemission spectroscopy. Characterizing the environmental stability of the two types of devices under illumination allows us to probe the role of ion induced electrode degradation on the stability of the devices in various atmospheres. We find that both types of devices show very similar degradation dynamics, revealing that suppressing ion induced electrode degradation does not improve the operational stability of these cells.

EXPERIMENTAL METHODS

**Device fabrication**

If not stated otherwise, all materials were purchased from Sigma Aldrich. To fabricate devices, pre-patterned ITO-coated glass substrates (PsiOTech Ltd., 15 $\Omega\text{sq}^{-1}$) were first cleaned sequentially with acetone and 2-propanol, followed by 10 min oxygen plasma treatment. PEDOT:PSS (Clevios Al 4083, Heraeus) was spin-coated onto the clean ITO substrates and then annealed at 150 °C for 10 min in air. For the perovskite layer fabricated by the lead acetate trihydrate recipe, $CH_3NH_3I$ (GreatCell Solar) and $Pb(Ac)_2 \cdot 3(H_2O)$ (3:1, molar ratio) were dissolved in anhydrous N,N-dimethylformamide (DMF) with a concentration of 40 wt% with the addition of hypophosphorous acid solution (6 µL mL$^{-1}$ DMF). The perovskite solution was spin-coated at 2000 rpm for 60 s in a drybox (RH < 0.5 %). After drying for 5 min, the samples were annealed at 100 °C for 5 min. Subsequently, the samples were transferred to a $N_2$ filled glovebox. PCBM (Solenne BV) in chlorobenzene (20 mg mL−1) was dynamically spin-coated at 2000 rpm for 45 s and annealed at 100 °C for 10 min. BCP was fully dissolved in 2-propanol (0.5 mg mL−1) and dynamically spin-coated at 4000 rpm for 30 s. ZnO nanoparticles were synthesized following an adapted procedure of Pacholski, which is briefly described in Supplementary Note 1.[28] The resulting nanoparticles were further characterized by UV-visible absorbance and transmission electron microscopy (TEM) (Figure S2 and Figure S3, Supporting Information). TEM characterization was carried out by a JEOL 2010 TEM (200 kV) equipped with a Gatan camera. Single or multiple layers of ZnO were cast from colloidal solution to obtain layers of different thicknesses. To complete the device, 80 nm silver electrodes were deposited via thermal evaporation under high vacuum.

**Device characterization and degradation**

To properly assess the degradation of the perovskite solar cells each device was identically prepared, stored in a nitrogen filled glovebox, and transferred to a sealed environmental box without exposure to ambient air. A constant flow of either nitrogen, nitrogen and oxygen (80:20, v:v) or humidified nitrogen (30% relative humidity) was connected to an environmental box. The oxygen percentage was controlled by adjusting the relative flow rate of $O_2$ to $N_2$, and monitored by a zirconia sensor (Cambridge Sensotet, Rapidox 2100) continuously before being connected to the environmental box. All of the devices were operated under simulated AM1.5 sunlight at 100 mW cm$^{-2}$ irradiance (Abet Sun 3000 Class AAA solar simulator) for 10 h (in open circuit condition), then "rested" in the dark for two hours before another two hour measurement period, bringing the total experiment time to 14 h. This rest period was performed in response to Nie *et al.*, who demonstrated that short rest periods could "heal" degraded devices.[14] The J–V measurements were performed with a Keithley 2450 Source Measure Unit. The cells were scanned from forward bias to short circuit and back at a rate of 0.5 V s$^{-1}$ after holding under illumination at 1.2 V for 2 s. The light intensity was calibrated with a Si reference cell (NIST traceable, VLSI) and corrected by measuring the spectral mismatch between the solar spectrum, the spectral response of the perovskite solar cell, and the reference cell. The mismatch factor was calculated to be approximately 11%.

**X-ray photoemission spectroscopy and depth profiling**

The perovskite devices investigated by photoemission spectroscopy (PES) measurements were fabricated and degraded as described above. The Ag electrodes were peeled off and transferred into an ultrahigh vacuum (UHV) chamber of the PES system (Thermo Scientific ESCALAB 250Xi) for XPS measurements. XPS measurements were performed using an XR6 monochromated Al Kα source (hν = 1486.6 eV) and a pass energy of 20 eV. Depth profiles were performed on the remaining layers of the devices using a MAGSIC Ar etching source.

RESULTS AND DISCUSSION

To investigate the role of ion induced electrode degradation on the operational stability of perovskite solar cells, we study the degradation of complete cells using methylammonium lead iodide ($MAPbI_3$) in the inverted ITO/PEDOT:PSS/$MAPbI_3$/PCBM/HBL/Ag architecture, as displayed in Figure 1. We fabricate two types of perovskite solar cells: one with the commonly used organic BCP HBL and another with ZnO nanoparticles with all other layers fabricated in an identical fashion. In order to provide the best possible comparison of the degradation behavior between devices with the two HBLs, our first step was to optimize the thickness of the ZnO layer in order to achieve a comparable initial photovoltaic performance of the two types of devices with similar charge extraction efficiency and recombination. ZnO nanoparticles were deposited *via* spin-coating at various spin speeds, and the resulting device's performance was measured under AM 1.5G solar illumination and are displayed in Figure 2. As the spin speed decreases (and thus the ZnO layer becomes thicker), device performance first increases up until approximately 8 nm of ZnO thickness and then decreases with further increasing ZnO thickness. At this thickness, both current and fill factor are maximized, while the voltage is largely independent of HBL thickness. At this optimal thickness the ZnO devices possess comparable performance to the BCP reference devices. Typical J-V curves for both device types are shown in Figure S1 (Supporting Information).

As mentioned before, previous work has suggested that one of the primary degradation mechanisms in perovskite-based devices is due to migration of ions towards and subsequent reaction with the metal electrodes. In the case of the devices in this study, this would be the migration of iodide to the silver electrode, which could then react with the silver to form the insulator AgI.[16,19,29] To examine whether the ZnO HBL did effectively block iodide migration, both ZnO and BCP devices were degraded in an inert environment and under constant solar illumination for 14 h, and afterwards the Ag electrodes were removed *via* tape and characterized by XPS. This allowed us to look for the presence of iodine in the electrodes. We note that since XPS experiments are performed in ultra-high vacuum, volatile iodine cannot be detected, so only species of bound iodide, such as in metal-iodides would be detected. The I3d signals for both the ZnO and BCP devices' electrodes are displayed in Figure 3a. While the BCP devices show a clear and strong I3d signal, indicating that iodide has indeed reacted with the electrode, the ZnO devices, in contrast, show no presence of iodine. This confirms that, on the timescale of the degradation experiment, a ZnO HBL effectively blocks the migration of iodide ions into a metal electrode, most likely due to the layer's higher density when compared to BCP. Figure 3b shows the excess iodine signal measured by XPS depth profiling, of degraded devices, from which the electrodes were peeled off. Similar to the results of Figure 3a, significant excess of iodine is detected in the PCBM extraction layer of the BCP device, while there is a far smaller iodine excess for the ZnO device. This result suggests that the inorganic HBL not only blocks iodide from reaching the metal electrode, but also prevents Ag from penetrating into the PCBM layer during the thermal evaporation of the electrode, where in turn it could also react with I ions. This mechanism has also been reported to be detrimental to the device performance.[30] These results confirm that the incorporation of an inorganic hole-blocking layer can suppress ion migration induced electrode degradation, while the organic HBL allowed for this degradation to take place.

Figure 4 displays the degradation behavior of the BCP devices under exposure to AM 1.5G solar illumination in the three different atmospheres: $N_2$, dry air, and humidified $N_2$ (30% RH) with an additional period at 10-12 hours where the cells were left to rest in the dark to investigate possible performance recovery.[14] We chose to compare three different degradation environments to eliminate the possibility that the difference in degradation dynamics would be associated with the different HBL, rather than the ion induced electrode degradation. Devices were held in the open circuit condition while not being measured. During the fourteen-hour measurement period we observe a moderate to severe decrease across all PV parameters, with $V_{OC}$ being the least affected and $J_{SC}$ the strongest. For the inert and dry air environments, $V_{OC}$ retains above 90% of its initial value after 14 hours. The devices exposed to water dipped below 90% of $V_{OC}$ after approximately 3 hours of measurement. $J_{SC}$ and FF plummet in all cases, falling to ~ 60% and 20% of their initial values for $N_2$/dry air and 30% RH respectively, and are the primary reasons for the steep loss in PCE. Interestingly, there appears to be little difference between the degradation behavior in inert and dry air environments, with devices exposed to humidity degrading markedly faster than both. Additionally, exposure to humidity appears to affect the various pixels on each device differently, as evidenced by the jagged trend lines and significantly wider error margins.

The degradation behavior for the experimental devices employing ZnO as HBL, under identical degradation conditions, is displayed in Figure 5. Overall, the results are very similar to the BCP devices, with a relatively stable $V_{OC}$ and strongly degraded $J_{SC}$ and FF. Just like the reference devices, the effects under $N_2$ and dry air atmospheres are identical, and exposure to water again degrades the devices both more quickly as well as less uniformly. The overall amount of degradation is slightly reduced for the ZnO devices, with the PCE falling to roughly 40% rather than 30% of their initial values. This is, however, only a slight improvement and could very easily be due to natural variation between devices.

Combining the degradation behavior with the XPS measurements of the electrodes yields two main conclusions. First, degradation for devices under constant illumination with the architecture ITO/PEDOT:PSS/CH$_3$NH$_3$PbI$_3$/PCBM/[BCP or ZnO]/Ag is not driven by an oxygen-related mechanism, as devices in both inert and dry air atmospheres displayed similar degradation characteristics. Aristidou *et al.* proposed that oxygen related perovskite decomposition proceeds by the generation of photoexcited electrons within the perovskite crystal.[3] Since PCBM is known to extract electrons from MAPbI$_3$ on shorter timescales than chemical reactions can occur,[31] the presence of PCBM at the perovskite interface possibly prevents oxygen-related degradation by withdrawing photoexcited electrons rapidly after their formation. The presence of moisture accelerates the degradation, regardless of the HBL used. With several vulnerable components, the root cause of the rapid decay in moisture is difficult to isolate. For example, while water is known to have a significant impact on perovskite stability[2,32–34] and also might facilitate the diffusion of volatile products inside the perovskite,[26,35] PCBM has also been shown to undergo strong, irreversible degradation in the presence of water.[36]

The second overall conclusion is that in contrast to the prevailing view, ZnO HBLs do not significantly alter the degradation characteristics when compared to the more common HBL BCP.[21] XPS measurements of degraded device electrodes did confirm that ZnO hinders the reaction of iodide ions with the electrode on the timescale of the experiment, however there was no meaningful difference between the ZnO and BCP device performance deterioration. Han *et al.*[37] found similar iodide infiltration into the silver electrode through the hole transport layer Spiro-OMeTAD in their devices, accompanied with a dramatic loss in PCE. They suggested that replacement of the silver contact with a more chemically inert electrode, such as Cu or C, could improve device stability by preventing this modification. Our results suggest that this iodide infiltration, and subsequent modification of the electrode into a more insulating species, cannot be the dominant factor in the observed overall performance loss. Lee *et al.* previously showed that the removal and re-evaporation of the silver electrode in degraded devices did partially restore the device's performance,[19] however a significant reduction in $J_{SC}$ remained. This suggests that an additional irreversible degradation mechanism is present.

In a study related to our work, Back *et al.* demonstrated that a layer of titanium suboxide (TiOx) placed between the PCBM and silver electrode could effectively increase device stability.[27] Indeed, on similar time scales (~10h) to our experiments, devices with a TiOx layer maintained ~ 80% of their initial PCE, compared to 50% with our ZnO HBL. This seemingly contradictory result can be explained by two factors. First, their reference devices contained no HBL on top of the PCBM. It is widely known that directly evaporating metallic contacts onto organic films can damage and thereby harm the film's stability.[38] Therefore, the TiOx layer will increase device stability simply by protecting the PCBM during thermal evaporation. Secondly, Back *et al.* held their devices at the maximum power point (MPP), which has been shown to slow degradation significantly when compared to devices held at $V_{OC}$.[39] Once this difference is accounted for, the stability for devices containing TiOx and ZnO nanoparticle blocking layers is extremely similar. In another study using a similar device structure, Akbulatov *et al*. demonstrated strongly enhanced stability when replacing PCBM with a perylene diimide derivative in encapsulated devices degraded in nitrogen under illumination. They attributed the strong degradation to the accumulation of volatile methylammonium iodide within the PCBM layer, subsequent reaction of iodide with the silver electrode, and the resulting formation of PbI$_2$ inside the perovskite layer.[20]

Taken all these results together, a plausible explanation for the observed deterioration of the device performance is obtained by considering the effects of ion migration not on the electrode, but instead on the perovskite active layer. While the ZnO layer prevents iodide infiltration into the electrode, the diffusion of volatile components out of the perovskite layer into PCBM should still be possible (and would remain undetectable by XPS) and has been previously shown to initiate significant degradation, e.g. by way of passivating the perovskite at the crystal grain boundaries.[20,26,29,40] Passivation at the boundaries in turn isolates each individual grain, leading to difficulties in conduction and charge extraction. This conclusion is supported by the specific mechanism of degradation observed in our study, with the primary drivers of PCE deterioration being a loss of $J_{SC}$ and FF, the two parameters associated with conductivity and charge extraction efficiency. Additionally, recently Zhao *et al* showed that ion induced degradation of charge extraction layers can also be a major cause for device instabilities.[41] Our results indicate that even without the iodide migration induced degradation of the electrode, the above mentioned degradation pathways can still take place and strongly affect device stability.

CONCLUSIONS

To summarize, we investigated the role of ion induced electrode degradation in inverted perovskite photovoltaic devices under realistic operational conditions. We find that contrary to what was previously suggested, suppressing degradation by blocking ion migration to the electrode does not necessarily improve device stability significantly under operational conditions. While it has been shown that this degradation mechanism plays a significant role in determining long-term dark storage stability of devices, its significance is diminished once more prominent degradation processes are present under full illumination. We propose that it is not ion-migration induced electrode corrosion, but rather the diffusion of volatile products out of the perovskite active layer and into the PCBM layer, that causes strong and irreversible degradation under operational conditions. Further research is required to elucidate the exact role that ion migration plays in affecting device performance and stability, however this work shows that the scope of such research must encompass a thorough examination, under realistic operating conditions, of both the perovskite active layer as well as the other device components.

FIGURES

(a) (b)

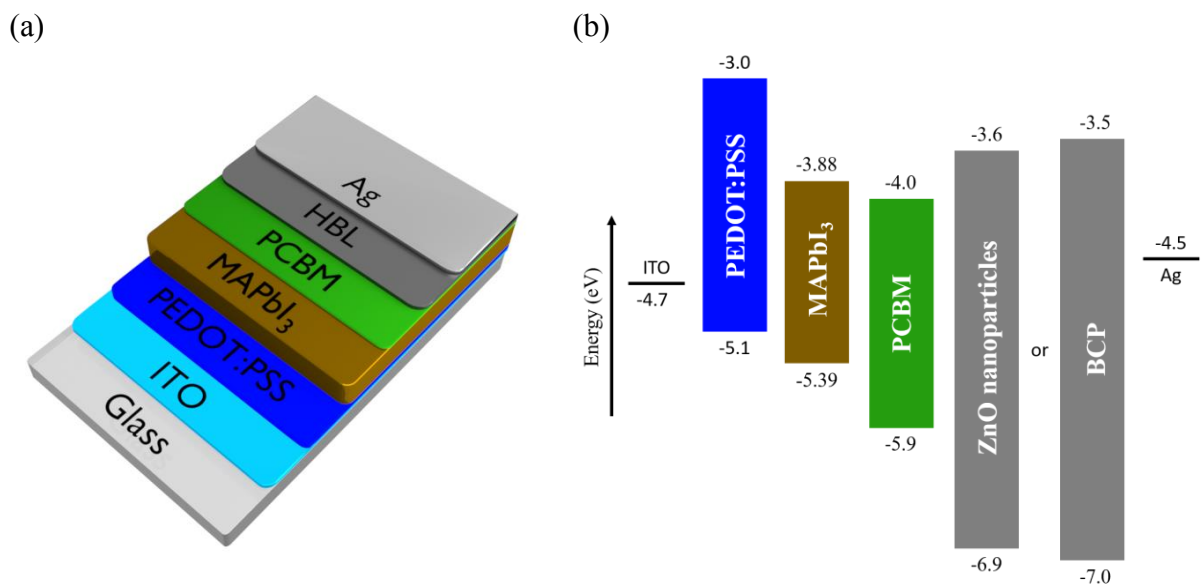

(c)

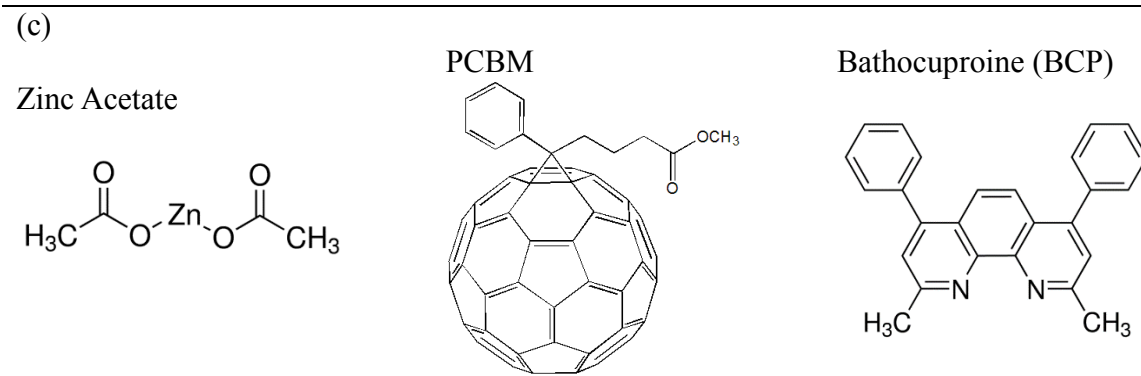

**Figure 1:** (a) Device architecture, (b) energy diagram, and (c) ETL/HBL materials used.

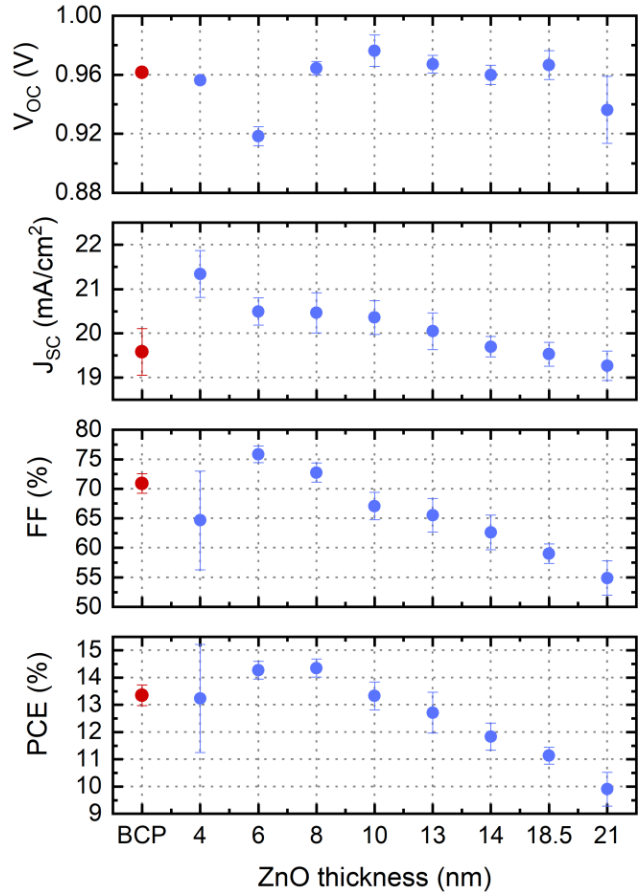

**Figure 2:** Photovoltaic (PV) parameters as a function of ZnO nanoparticle layer thickness. One sample with BCP as HBL was fabricated in the same batch for direct comparison. Error bars represent the standard deviation. An optimal value of 8 nm was found for ZnO nanoparticle layer's thickness.

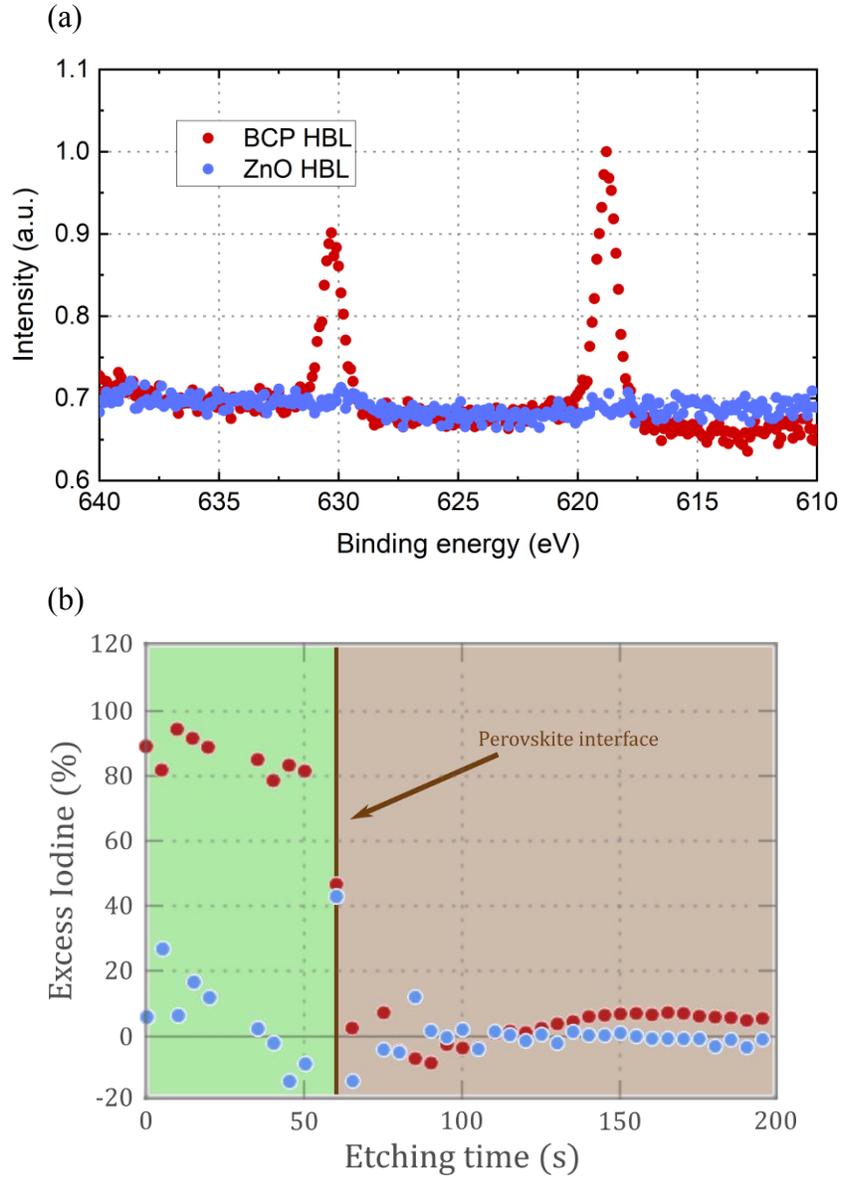

**Figure 3:** (a) XPS measurement of the surface of the Ag electrodes of fully degraded devices, showing the I3d (iodine) signal. (b) Excess Iodine in the Ag/HBL/PCBM layer, obtained by XPS conducted after etching via argon beam. After approximately 60s of etching the interface with the perovskite film is reached.

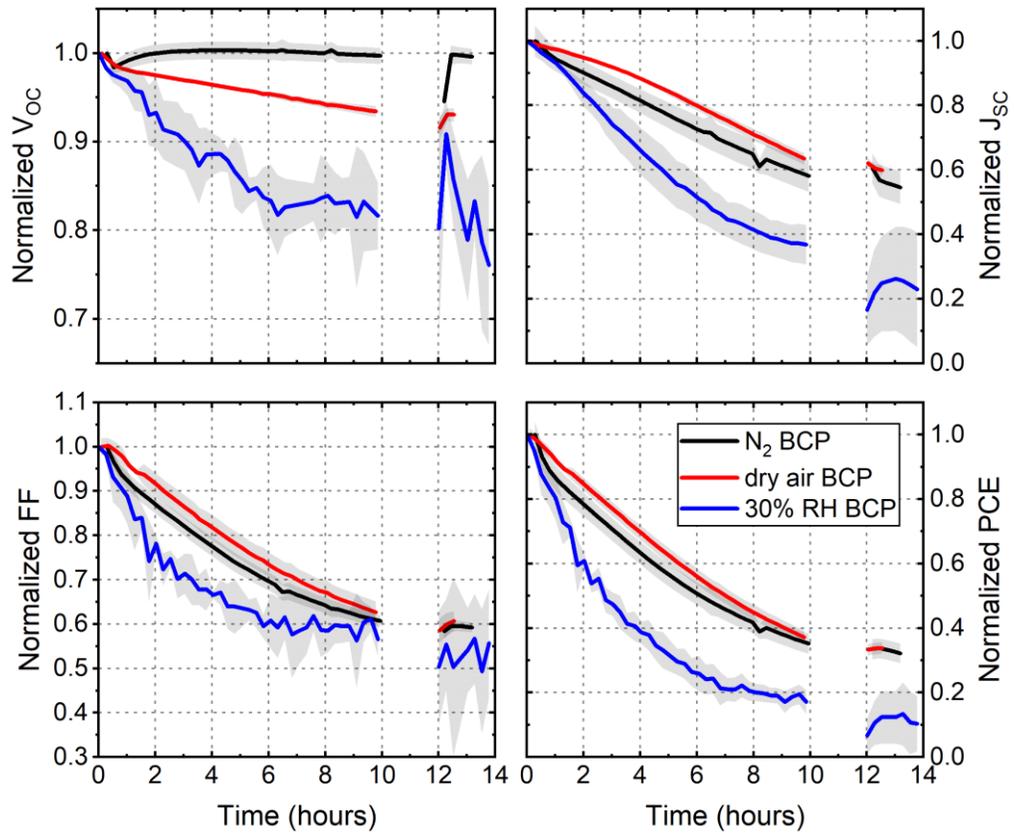

**Figure 4:** Evolution of PV parameters (open-circuit voltage, short-circuit current, efficiency, and fill factor) for devices with BCP as HBL, degraded in N$_2$, dry air, and humidified N$_2$ (30% RH) atmospheres over 14 hours. From hours 10-12 the devices were left in the dark, in order to test the reversibility of the degradation. The shaded region represents the standard deviation for the measurement.

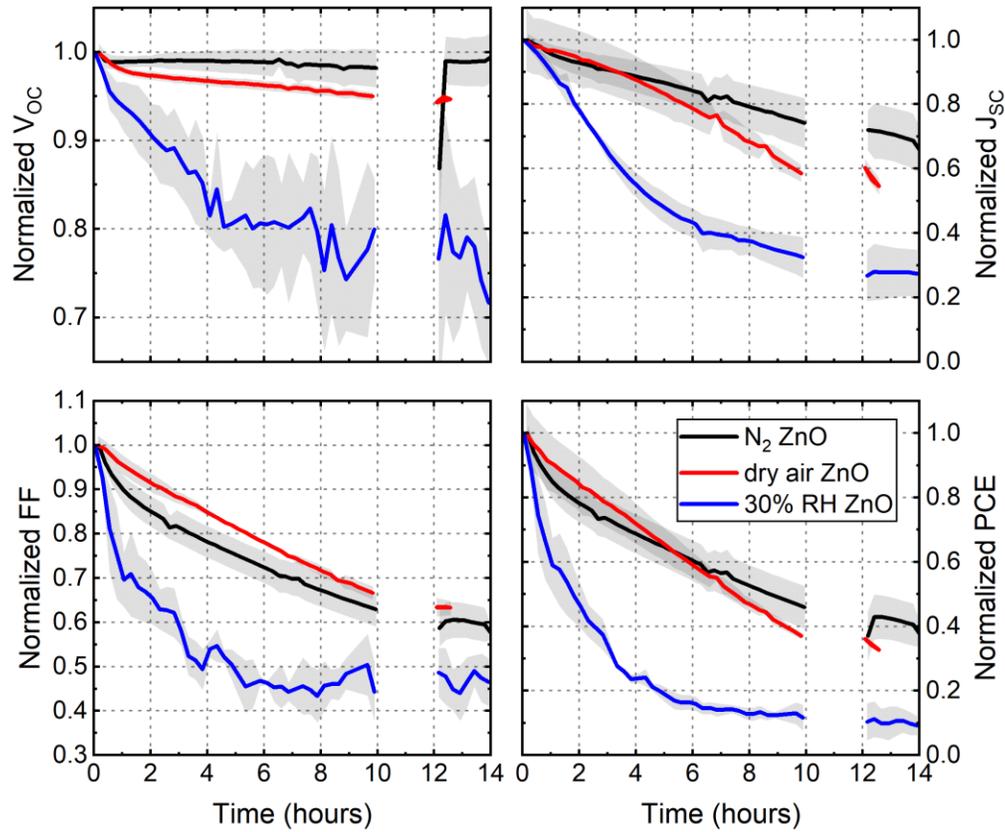

**Figure 5:** Evolution of PV parameters (open-circuit voltage, short-circuit current, efficiency, and fill factor) for devices with ZnO as HBL, degraded in $N_2$, dry air, and humidified $N_2$ (30% RH) atmospheres over 14 hours. From hours 10-12 the devices were left in the dark, in order to test the reversibility of the degradation. The shaded region represents the standard deviation for the measurement.

## ASSOCIATED CONTENT

**Supporting Information**. Synthesis and characterization of ZnO nanoparticles, J-V curves of fresh photovoltaic devices

## AUTHOR INFORMATION

**Corresponding Author**

*email: vaynzof@uni-heidelberg.de

Kirchhoff Institute for Physics and Centre for Advanced Materials, Im Neuenheimer Feld 227/225, 69120, Germany

**Author Contributions**

The manuscript was written through contributions of all authors. All authors have given approval to the final version of the manuscript. This work has received funding from the European Research Council (ERC) under the European Union's Horizon 2020 research and innovation programme (ERC Grant Agreement n° 714067, ENERGYMAPS).

ACKNOWLEDGMENT

The authors would like to kindly thank Prof. Uwe Bunz for providing access to the device fabrication facilities. P.F. thanks the HGSFP for scholarship.